\documentclass[twocolumn,showpacs,superscriptaddress,preprintnumbers,amsmath,amssymb]{revtex4}

\usepackage{graphicx}
\usepackage[usenames]{color}

\begin{document}

\title{Anderson localization of partially-incoherent light}

\author{D.~\v{C}apeta}
\affiliation{Department of Physics, University of Zagreb, PP 332, 10000 Zagreb, Croatia}
\author{J.~Radi\'{c}}
\affiliation{Department of Physics, University of Zagreb, PP 332, 10000 Zagreb, Croatia}
\author{A.~Szameit}
\affiliation{Technion, Israel Institute of Technology, Haifa, Israel}
\author{M.~Segev}
\affiliation{Technion, Israel Institute of Technology, Haifa, Israel}
\author{H.~Buljan}
\affiliation{Department of Physics, University of Zagreb, PP 332, 10000 Zagreb, Croatia}


\date{\today}

\begin{abstract}
We study Anderson localization and propagation of partially-spatially incoherent 
wavepackets in linear disordered potentials, motivated by the insight that interference phenomena 
resulting from multiple scattering are affected by the coherence of the waves. 
We find that localization is delayed by incoherence: the more incoherent the 
waves are, the longer they diffusively spread while propagating in the medium. 
However, if all the eigenmodes of the system are exponentially localized 
(as in one- and two-dimensional disordered systems), any partially-incoherent wavepacket 
eventually exhibits localization with exponentially-decaying tails, after 
sufficiently long propagation distances. 
Interestingly, we find that the asymptotic behavior of the incoherent beam is 
similar to that of a single instantaneous coherent realization of the beam.
\end{abstract}

\pacs{42.25.Dd,42.25.Kb,72.15.Rn}
\maketitle


The phenomenon of Anderson localization was conceived in 
the context of disordered electronic systems \cite{Anderson1958}, however, 
localization phenomena have been extensively studied also in other systems 
\cite{John1984,Anderson1985,Akkermans1985,Albada1985,Wolf1985,Raedt1989,Wiersma1997,
Chabanov2000,Storzer2006,Schwartz2007,Lahini2008,Szameit2010,Lagendijk2009,Billy2008,Roati2008,
Sanchez-Palencia2010} including optics \cite{John1984,Anderson1985,Albada1985,
Wolf1985,Raedt1989,Wiersma1997,Chabanov2000,Storzer2006,Schwartz2007,Lahini2008,Szameit2010,Lagendijk2009} 
and ultracold quantum gases \cite{Billy2008,Roati2008,Sanchez-Palencia2010}. 
For a direct observation of the phenomena, optical and ultracold atomic 
systems have some profound advantages over the condensed matter systems:
the influence of the environment such as thermal fluctuations and phonons can be minimized to 
become negligible, and nonlinearity/interactions can be controlled and virtually turned off.
Nonlinearity in optics can be controlled by the intensity of light, whereas 
interactions of ultracold gases can be tuned via Feshbach resonances. 
In contrast, electron-electron and electron-phonon interactions are always 
present in condensed matter systems. 
These influences are important since they affect interference, and Anderson 
localization arises from interference among multiple scattering events from 
disorder in the medium.

Here we address Anderson localization of waves with imperfect 
coherence. We investigate whether an initial finite-size partially-incoherent 
wavepacket would spread through a linear disordered potential, or would its spreading 
be stopped by virtue of disorder?  
We demonstrate our findings on an optical (1+1)D potential, and 
discuss the implications of our results to other wave systems in nature.
A superficial answer to the question above 
may be that, since incoherence destroys interference effects, a sufficiently 
incoherent wavepacket will diffuse through the random medium without ever being localized. 
However, from the theory of coherence \cite{MandelWolf} it is known that an 
incoherent wave can be thought of as a superposition of coherent modes with 
stochastically varying coefficients. Each of these coherent modes is expected 
to undergo localization, hence the entire wavepacket should localize. Our study 
reveals that coherence indeed affects the properties of evolving wavepackets, 
in a sense that more incoherent beams spread more while propagating through
the random medium. However, if the system's eigenmodes are all localized, as 
is the case for all one-dimensional [(1+1)D] and two-dimensional [(2+1)D] fully 
disordered systems \cite{Economou}, the incoherent wavepacket will eventually 
become Anderson localized. 
Finally, we find that a typical instantaneous coherent speckled realization 
of the incoherent beam exhibits the same asymptotic behavior as the time-averaged 
incoherent wavepacket.

The idea that localization could be observed in optics dates back to the 
beginning of the 80s \cite{John1984,Anderson1985}. The experiments on so-called 
weak localization \cite{Akkermans1985}, which can be pictured 
in terms of a coherent backscattering process, were soon to follow \cite{Albada1985,Wolf1985}. 
Experiments on strong localization in random media were performed in various systems 
\cite{Wiersma1997,Chabanov2000,Storzer2006}. In 1989, a non-traditional 
idea for observing localization was proposed: the transverse localization 
scheme \cite{Raedt1989}, which exploits the equivalence between the Schr\" odinger 
equation and the paraxial wave equation for light. Indeed, this scheme was used 
for a clear demonstration of Anderson localization in random optical lattices 
\cite{Schwartz2007}, for the observation of Anderson modes \cite{Lahini2008}, 
and localization near an interface \cite{Szameit2010}. 
All of these have dealt with fully coherent waves only. 
In a different domain, the propagation of partially-incoherent light in random 
media is a subject of considerable interest (e.g., see \cite{Gbur2002}). However, 
in the context of localization, the only studies on incoherent light were on 
enhanced backscattering \cite{Wax2000,Kim2004}, which is considered a precursor to 
Anderson localization. To the best of our knowledge, strong localization with 
partially-incoherent waves has never been studied.

Consider the propagation of a partially-spatially incoherent 
optical beam, linearly polarized, originating from a quasimonochromatic 
continuous-wave source. 
Such a beam can be constructed by sending a laser beam through a rotating 
diffuser (e.g., see \cite{Mitchell1996}). 
Its state at a given propagation distance $z$ can be described 
in terms of the mutual coherence function \cite{MandelWolf}, 
\begin{equation}
B(x_1,x_2,z)=\langle E^*(x_2,z,t) E(x_1,z,t) \rangle_t,
\label{MCF}
\end{equation}
where $\langle \cdots \rangle_t$ is the time-average, and $E$ is the stochastic field. 
Instead of $B(x_1,x_2,z)$, the state of the system can be described by 
an orthonormal set of modes $\psi_j(x,z)$ ($j=1,2,\ldots$), and 
their modal weights $\lambda_j$, which are obtained from 
the eigenvalue equation \cite{MandelWolf}
\begin{equation}
\int dx_2 B(x_1,x_2,z) \psi_j(x_2,z) = \lambda_j \psi_j(x_1,z),
\end{equation}
i.e., $B(x_1,x_2,z)=\sum_j \lambda_j \psi_j^*(x_2,z)\psi_j(x_1,z)$. 

We analyze linear propagation of such a beam in the transverse localization 
scheme \cite{Raedt1989}, in a waveguide array defined by 
$n^2=n_0^2+2 n_0 \delta n (x)$, where $n_0$ is a constant term, while 
$\delta n (x)$ describes disorder. 
The propagation of the beam along the $z$ axis is governed by the 
Schr\" odinger equation \cite{Schwartz2007}
\begin{equation}
i\frac{\partial\psi}{\partial z}=
-\frac{1}{2k} \frac{\partial^2 \psi}{\partial x^2}
-\frac{\delta n(x) k }{n_0} \psi,
\label{evolOpt}
\end{equation}
where $k=n_0\omega/c$ is the wave vector, $\omega$ is the temporal 
frequency of the beam carrier, and $c$ is the speed of light. 

In our simulations, we analyze the evolution of Gaussian input beams:
\begin{equation}
B(x_1,x_2,z=0)=I_0 \exp[- \left( \frac{x_1+x_2}{2\sigma_I} \right)^2
- \left( \frac{x_1-x_2}{\sigma_C} \right)^2],
\label{B_Gauss}
\end{equation}
where $\sigma_I$ and $\sigma_C$ are the spatial and the coherence widths 
of the beam, respectively. 
In disordered media, the propagation depends on the particular realization of 
the random potential $\delta n (x)$. Thus, to obtain meaningful results, one 
needs to observe the evolution of the mutual coherence for many realizations 
of the disorder, and calculate the disorder ensemble average \cite{Schwartz2007}: 
$\langle B(x_1,x_2,z) \rangle_d$.

\begin{figure}
\centerline{
\mbox{\includegraphics[width=0.4\textwidth]{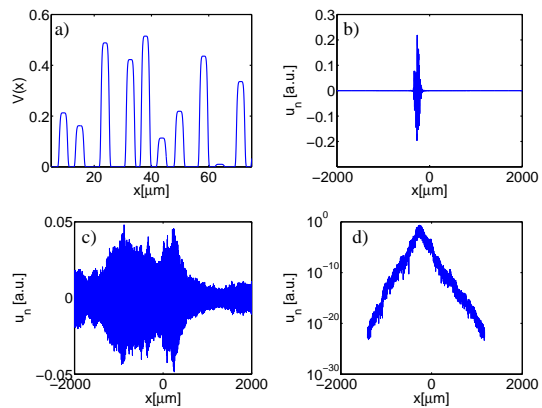}}
}
\caption{
The disordered potential and its eigenmodes. 
(a) Small section of the dimensionless disordered potential $V(x)=-2 \delta n(x) (kx_0)^2/n_0$, 
where $k=2\pi n_0/\lambda$, $\lambda=514$~nm, $x_0=1$~$\mu$m, $n_0=1.45$. 
(b,c) Typical eigenfunctions (Anderson states) of the structure with a smaller (b) 
and larger (c) spatial extent, in a sample of width $L=4$~mm. 
(d) Logarithmic plot of $|u_n(x)|$ from (b) with exponentially decaying tails, 
which are the fingerprint of Anderson localization. 
}
\label{delta_n}
\end{figure}

The disordered potential used in our simulations is illustrated in Fig. \ref{delta_n}(a). 
The index of refraction varies randomly (with uniform distribution) 
between $\delta n=0$ and $\delta n=1.2 \times 10^{-3}$. 
The width of every rectangular potential unit shown in Fig. \ref{delta_n}(a) 
is $2.7$~$\mu$m, but their mutual distances are random: 
first we fix the leftmost rectangle in its position, and then add the adjacent ones 
to the right such that their distance (center to center) is between $5$ and $9$~$\mu$m 
(chosen at random), and so on. 
Such a disordered medium can be created experimentally by using the ultrafast direct 
laser writing technique \cite{Szameit2010}.

Because the system is linear, the evolution of the beam (in a given realization 
of the disorder) can be described in terms of the modes of the system $u_n$ and 
their propagation constants $\beta_n$, which obey 
\begin{equation}
\beta_n u_n=
\frac{1}{2k} u_n''(x)
+\frac{\delta n(x) k }{n_0} u_n.
\label{eigen0}
\end{equation}
Every initial coherent wave $\psi_j(x,0)$ is projected onto the system's modes 
$u_n$: $c_{j,n}= \int dx \psi_j(x,0) u_n^*(x)$, yielding 
$\psi_j(x,z)=\sum_n c_{j,n} u_n(x) e^{i\beta_n z}$, that is,
\begin{equation}
B(x_1,x_2,z)=\sum_{j,n,m} \lambda_j c_{j,n}^* c_{j,m} u_n^*(x_2) u_m(x_1) e^{i(\beta_m-\beta_n) z}.
\label{Bgeneral}
\end{equation}
It is reasonable to conjecture that after sufficiently long propagation $z$, 
the beam will dephase and attain the following form 
of the mutual coherence:
\begin{equation}
B(x_1,x_2,z\rightarrow \infty)=\sum_{j,n} \lambda_j |c_{j,n}|^2 u_n^*(x_2) u_n(x_1).
\label{Bfinal}
\end{equation}
From Eqs. (\ref{Bgeneral}) and (\ref{Bfinal}) we clearly see that, if the medium is 
infinitely broad, and if the potential is fully random, then, since all the 
eigenmodes $u_n$ are Anderson localized, any initial finite-width partially-incoherent beam will 
not diffuse during propagation despite its incoherence. 
One can expect that, after sufficiently long propagation, the tails of the incoherent 
beam will become exponentially decaying: $B(x,x,z\rightarrow \infty)\propto \exp (- \gamma |x|)$, 
with $\gamma$ corresponding to the excited eigenmode $u_n$ which has the slowest 
decay. 
However, in reality, all samples have finite transverse ($L$) and 
longitudinal ($Z$) size; hence, we explore the finite-size effects.

\begin{figure}
\centerline{
\mbox{\includegraphics[width=0.4\textwidth]{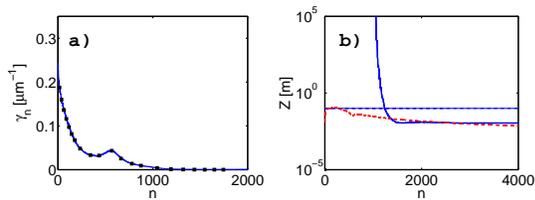}}
}
\caption{
(a) Lyapunov exponents $\gamma_n$ extracted from the eigenfunctions $u_n$ (solid 
blue line) and via the transfer matrix approach (black squares). 
Shown are averages over 40 different realizations of the random potential; 
the standard deviation is smaller than the thickness of the line. 
(b) The Thouless (solid blue line) and the Heisenberg (dashed red line) 
times (propagation distances) vs. $n$ for the system with $L=4$~mm.
Horizontal line depicts $Z=10$~cm. See text for details. 
}
\label{LE}
\end{figure}

First we calculate the eigenvectors/eigenvalues of the potential $\delta n(x)$. 
We use the LAPACK implementation of the MRRR algorithm \cite{Dhillon1997},
with Dirichlet boundary conditions $u_n=0$ at the 
boundaries of the medium ($x=\pm L/2$).
Typical eigenfunctions are shown in Figs. \ref{delta_n}(b)-(d). 
Up to some critical value of $n$, the spatial extent of the eigenfunctions 
$u_n$ is smaller than $L$ [see Fig. \ref{delta_n}(b)], whereas above this 
value they extend to one or both of the boundaries [see Fig. \ref{delta_n}(c)]. 
The localized eigenfunctions are a consequence of Anderson localization, which can be seen 
from Fig. \ref{delta_n}(d), where we depict 
$|u_n(x)|$ on a logarithmic scale. Clearly, the amplitude of $u_n(x)$ decays 
exponentially, $\log |u_n(x)| \propto \exp(\gamma_n x)$, which is a fingerprint of 
Anderson localization in this (1+1)D system. 
From these eigenmodes, we calculate their Lyapunov exponents as follows: 
We form a set $\{ |u_n(x_{\textrm{max}}) | \}$ of local maxima of $|u_n(x)|$, 
and then fit $-\gamma_n |x-x_c|+b$ to the set of points 
$\log |u_n(x_{\textrm{max}})|$, to obtain $\gamma_n$ for every $u_n$. 
For the lowest eigenstates, which have too few local extrema, $|u_n|$ was fitted. 
The Lyapunov exponents, averaged over 40 different realizations of the random 
potential, are shown in Fig. \ref{LE}(a) as a function of $n$; 
the standard deviation is as small as the thickness of the line implying that 
the Lyapunov exponents have the self-averaging property. In order to underpin 
our results, we have also calculated the Lyapunov exponents via the transfer matrix 
approach \cite{Economou}, i.e., for an essentially infinite sample, and obtained 
nondistinguishable results [black squares in Fig. \ref{LE}(a)]. 
Evidently, the overall trend is such that the Lyapunov exponents decrease 
with increasing $n$ (decreasing $\beta_n$), except for a small hump, which is a consequence of 
relatively small fluctuations in distances between adjacent potential peaks [Fig. \ref{delta_n}(a)]
(if we allow larger fluctuations the hump disappears). 

\begin{figure}
\centerline{
\mbox{\includegraphics[width=0.4\textwidth]{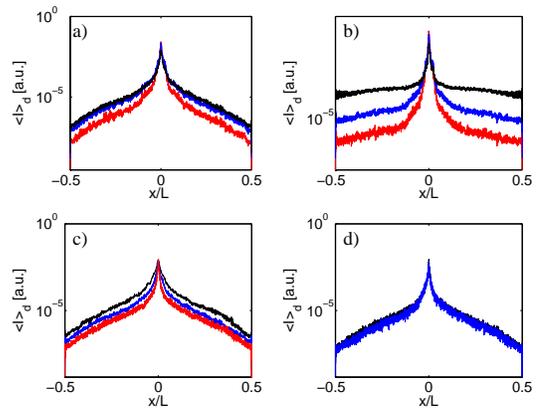}}
}
\caption{
Intensities of the incoherent beam 
after $Z=10$~cm of propagation, for different system parameters,
averaged over 40 realizations of the disorder. 
(a,b) Beam intensity for absorbing (a) and reflecting (b) boundary conditions for beams with 
$\sigma_I=10$~$\mu$m, and $\sigma_C=1$~$\mu$m (top black line), $3$~$\mu$m (middle blue line), 
and $5$~$\mu$m (bottom red line); $L=4$~mm.
(c) Beam intensity for absorbing boundary conditions and three values of 
$L$: $2$, $4$, and $8$~mm (top to bottom) [$\sigma_I=10$~$\mu$m, $\sigma_C=1$~$\mu$m].
(d) The intensity for one specific instantaneous initial realization of the 
incoherent field (lower blue line), and the time-averaged intensity 
(upper black line) [$\sigma_I=10$~$\mu$m, $\sigma_C=1$~$\mu$m, $L=8$~mm, 
absorbing boundary conditions].
}
\label{inten}
\end{figure}

In a realistic experimental situation with finite-size samples, 
a partially-incoherent initial wavepacket is likely to excite both 
types of modes: modes that are Anderson localized on a scale smaller than $L$, 
and modes extending to the boundary. In order to investigate the 
finite-size effects, we first evolve (numerically) an initial wavepacket with 
absorbing boundary conditions [Figure \ref{inten} (a)], and then repeat the 
simulation with reflecting boundary conditions [Figure \ref{inten} (b)] 
(both can be realized in experiments). 
The initial size of the beam is $\sigma_I=10$~$\mu$m, and the degree of 
coherence varies: $\sigma_C=1,3$, and $5$~$\mu$m; $L=4$~mm. 
In our simulations, absorbing boundary conditions are included as an imaginary 
index of refraction, $\Im \delta n_{\textrm{abs}}>0$, which is present {\em only} 
close to the sample edges: $\delta n_{\textrm{abs}}=0$ for $|x|< 0.96 \frac{L}{2}$.

From the simulations with absorbing edges we find that, after sufficiently long 
propagation, the intensity structure has exponentially decaying tails: 
$\langle I \rangle_d\propto \exp(-\gamma |x|)$. All graphs have approximately the 
same value for the slope, $\gamma=3.2\times 10^{-3}$~$\mu$m$^{-1}$. 
This is in accordance with our analysis of eigenstates. Namely, the part of the 
beam exciting eigenstates which touch (or are in the very vicinity of) 
the edges, gets absorbed during propagation. The remaining part of the 
beam excites only exponentially decaying eigenstates, which yields 
the intensity plotted in Fig. \ref{inten}(a). 
The value of $\gamma$ found from graphs in Fig. \ref{inten}(a) 
correspond to the slowest decay rate $\gamma_n$ of the eigenstate $u_n$ 
which is excited, and which does not overlap with the absorbing boundary,
i.e., $\langle u_n | \delta n_{\textrm{abs}}(x) | u_n \rangle\approx 0$
(one should take into account factor 2 because $I=\langle|E|^2\rangle_t$).

At this point it is worthy to consider the Thouless ($Z_{T}$) and Heisenberg ($Z_{H}$) 
times (propagation distances) in our system. The former tells us the 
average time it takes for a particle (with some energy) to diffuse across 
the sample, and the latter is the longest time the particle can 
travel inside the sample without visiting the same region twice 
(e.g., see \cite{Lagendijk2009} and Refs. therein). 
In our case $Z_{T}$ corresponds to the inverse linewidth ($1/\Delta\beta_n$) 
of the eigenstates with absorbing boundary conditions. 
In calculating $Z_{T}(n)$ we have averaged $1/\Delta\beta_n$ 
over 20 adjacent eigenstates, and then over 40 realizations of the potential. 
In the same fashion we have calculated the average inverse level spacing 
which yields the Heisenberg time $Z_{H}(n)$.
Figure \ref{LE}(b) shows $Z_{T}$ and $Z_{H}$ vs. $n$. 
Evidently the Thouless time is effectively infinite for $n<1200$ indicating 
Anderson localization in our finite sample, whereas for greater $n$ values the 
two times are on comparable scales. 
From Fig. \ref{inten}(a), we conclude that the more coherent the light is - 
the stronger the localization effect is, in a sense that less power is 
absorbed by the walls. For the samples used in our simulations the rate of 
decaying tails are identical for all values of $\sigma_C$, 
because in every case the highest of the localized modes was excited.

It is interesting to note that the values $\gamma$ obtained after 
transport simulations (expansion) {\it depend} on the size of the sample $L$, 
and scale approximately as $\gamma \propto L^{-1}$. 
We checked this trend for $L=2,4$, and $8$~mm, see Fig. \ref{inten}(c). 
The observed scaling is found to be in agreement with the studies conducted 
in \cite{Anderson1980}, where it was shown that the logarithm of the 
dimensionless resistance scales as the length in case of localization.

\begin{figure}
\centerline{
\mbox{\includegraphics[width=0.3\textwidth]{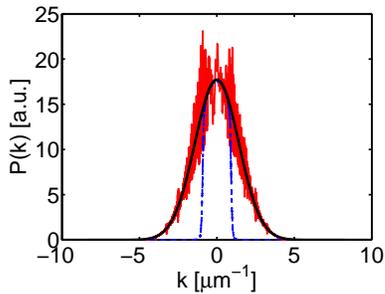}}
}
\caption{
Spatial power spectrum of the incoherent beam after $Z=10$~cm of propagation in a 
disordered medium with absorbing (thin dashed blue line) and 
reflecting (solid red line) edges, for an initial beam (thick solid black line) 
defined by $\sigma_I=10$~$\mu$m, $\sigma_C=1$~$\mu$m. All plots are 
averages over 40 realizations of the disorder. 
}
\label{k-abs}
\end{figure}

Figure \ref{k-abs} shows a typical change in the spatial power spectrum 
(closely related to the spatial coherence) for propagation 
dynamics with different boundary conditions. In the case of absorbing edges, 
there is a clear cut off in $k$-space above which the modes are no longer 
localized on a scale smaller than $L$. 
The modes $u_n$ comprised of plane waves (via Fourier transform) with $k$ 
values above the cut-off evidently spread to the absorbing 
boundaries of the medium. This process also increases the overall 
coherence of the beam, because it has fewer and fewer modes as it propagates. 
In contrast, for reflecting boundary conditions, 
the final power spectrum has the same overall shape as the initial spectrum, with fluctuations 
on top of the average shape.

A partially-spatially incoherent (quasimonochromatic) beam can be thought of as 
a train of fully coherent fields with intricate field structures, which are replaced 
on a time scale $\sim \Delta t_c$ (the temporal coherence time); quasimonochromatic 
means that $\Delta t_c \gg Z/c$. 
In experiments, one instantaneous realization of the incoherent field
can be observed by simply stopping the rotation of the diffuser \cite{Mitchell1996}. 
Interestingly, we find that one instantaneous field realization 
will asymptotically have exponentially-decaying tails similar to those of the 
time-averaged intensity, as illustrated in Figure \ref{inten}(d) 
(both graphs are averaged over disorder).
Thus, the localization of a partially-incoherent field can be thought of 
as the localization of many instantaneous coherent speckled structures, 
whose ensemble average produces localized mutual coherence. 
The exponential decay of the asymptotic intensity evidently depends both on 
the degree of coherence and the spatial structure of the input field. 

We can extend our conclusions to hold for spatially and 
temporally incoherent beams of light ($\Delta t_c<Z/c$), such as the one 
constructed from an incandescent light bulb in Ref. \cite{Mitchell1997}. 
In this case, a space-time mutual coherence function describing the beam 
can be transformed into the space-frequency domain \cite{MandelWolf,Buljan2002}. 
Because our results hold for every frequency component separately (with 
somewhat different numerical values because different frequencies experience 
different diffraction), due to linear superposition, it is evident that they 
also hold for the whole spatially and temporally incoherent beam. 

Before closing, we note that our results correspond to partially 
condensed {\it noninteracting} Bose-Einstein condensates (BECs) in 1D 
disordered potentials. 
As an interesting consequence of dimensionality, 
we note that our results also pertain to the strongly interacting 1D 
Bose gases \cite{Radic2010}, and noninteracting 1D Fermi gases, 
which are modeled by identical equations \cite{Buljan2006}. 
To mimic partially-condensed interacting BECs \cite{Buljan2005}, one should 
include nonlinearity, which we leave for further studies. 

In conclusion, we have predicted the properties of Anderson localization of 
partially spatially-incoherent beams in disordered linear (1+1)D photonic structures. 
We conclude that more incoherent light will diffusively spread more 
through the random medium than coherent light, however, the incoherent wavepacket 
will display exponentially decaying tails after sufficiently long propagation. 
We have discussed the finite size effects, and extended our conclusions 
for spatially and temporally incoherent light beams. 

This work was supported the Croatian-Israeli scientific cooperation program
funded by the Ministries of Science of the State of Israel and the Republic of Croatia.
AS thanks the Leopoldina - The German Academy of Science (grant LPDS 2009-13), and 
HB the MZO\v{S} grant 119-0000000-1015.


\end{document}